\documentclass[prl,superscriptaddress,amsfonts,showpacs,twocolumn]{revtex4}
\usepackage{times}
\usepackage{bm}
\usepackage{float}
\usepackage{amsfonts}
\usepackage{graphicx}
\usepackage[T1]{fontenc}
\usepackage{color}

\begin{document}

\title{Runaway electrification of friable self-replicating granular matter}

\author{Julyan H. E. Cartwright}
\affiliation{Instituto Andaluz de Ciencias de la Tierra (IACT), CSIC--Universidad de Granada, E-18100 Armilla, Granada, Spain}
\author{Bruno Escribano}
\affiliation{Basque Center for Applied Mathematics (BCAM), Alameda de Mazarredo 14, E-48009 Bilbao, Basque Country, Spain}
\author{Hinrich Grothe}
\affiliation{Institut f\"ur Materialchemie, Technische Universit\"at Wien, Getreidemarkt 9/BC/01, A-1060 Wien, Austria}
\author{Oreste Piro}
\affiliation{Departament de F\'isica, Universitat de les Illes Balears, E-07122 Palma de Mallorca, Spain}
\author{C. Ignacio Sainz D\'{\i}az}
\affiliation{Instituto Andaluz de Ciencias de la Tierra (IACT), CSIC--Universidad de Granada, E-18100 Armilla, Granada, Spain}
\author{Idan Tuval}
\affiliation{Institut Mediterrani d'Estudis Advan\c ats (IMEDEA), CSIC--Universitat de les Illes Balears, E-07190 Mallorca, Spain}

\pacs{45.70.Mg  05.45.-a  45.70.Qj  83.10.Pp}

\date{version of \today}

\begin{abstract}
We establish that the nonlinear dynamics of collisions between particles favors the charging of a insulating, friable, self-replicating granular material that undergoes nucleation, growth, and fission processes; we demonstrate with a minimal dynamical model that secondary nucleation produces a positive feedback in an electrification mechanism that leads to runaway charging. We discuss ice as an example of such a self-replicating granular material: We confirm with laboratory experiments in which we grow ice from the vapor phase in situ within an environmental scanning electron microscope that 
charging causes fast-growing and easily breakable palm-like structures to form, which when broken off may form secondary nuclei. We propose that thunderstorms, both terrestrial and on other planets, and lightning in the solar nebula are instances of such runaway charging arising from this nonlinear dynamics in self-replicating granular matter. 
\end{abstract}

\maketitle

Charging of grains of identical insulating materials during collisions has been of considerable interest recently, both for its intrinsic physics and for its applications to situations ranging from volcanic dust plumes and desert sandstorms to industrial powder processing \cite{kok2009,pahtz}.  However, work up to now has not explained why in some instances charging grows --- rather than diminishes as one might naively expect --- and can run away extremely rapidly, leading to electrical discharges: lightning. Outstanding examples of runaway collisional charging involve ice, in thunderstorms both on Earth \cite{vonnegut} and on other planets \cite{aplin,yair} and, it is speculated, in the solar nebula \cite{gibbard,pilipp,muranushi};  these instances concern granular media whose particles also undergo nucleation, growth, and fission, so that they, in effect, reproduce. In this work we demonstrate with a minimal dynamical model that secondary nucleation --- production of a new particle from an existing particle --- is a key process in producing runaway electrical charging in self-replicating granular matter. We concomitantly present the results of experiments on growing ice in situ from the vapour phase within an environmental scanning electron microscope. We show that an effect of the electric field is to induce the formation of fast-growing ice `palms' intermediate in morphology between whiskers and dendrites; the ease of breakage of these palm-like formations will clearly favor secondary nucleation and hence runaway electrification.

\begin{figure}[b]
\includegraphics*[width=0.8\columnwidth,clip=true]{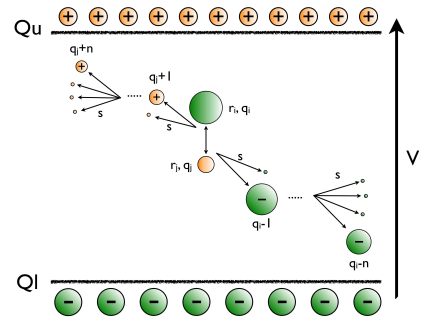}
\caption{\label{model}
Anatomy of our minimal model of charging of friable self-replicating granular matter  through secondary nucleation,
incorporating the processes of particle growth ($r_i \rightarrow r_i +1$), advection (with speed $u_i$), collision, charge transfer, and fission (with probability $s$). }
\end{figure}

``In a thousand seconds, more or less, its volume increases a thousandfold, the intensity of its electric fields increases a thousandfold, and its electric energy increases a billionfold''; thus described Vonnegut the tremendous metamorphosis a cumulus cloud undergoes to become a cumulonimbus or thundercloud \cite{vonnegut}. Thundercloud electrification is a consequence of ice particles colliding within a cloud and exchanging electrical charge. 
 Charge dipole development in a thunderstorm is due to the physical separation of particles with opposite charges inside the cloud \cite{springer_chapter}: Larger, heavier particles will fall, while smaller, lighter particles will rise in the updraft, and these particles carry different charges \cite{mason}.
Usually in thunderstorms the smaller ice particle is an ice crystal and carries positive charge aloft in the updraft; the larger graupel ice particle falls with an opposite negative charge, leading to a typical thunderstorm. We may contrast the foregoing with charging in other granular media \cite{Shinbrot2008,lacks}, where it has been argued that simple geometry, without growth and fission processes, leads to a net transfer of electrons from larger to smaller particles \cite{kok2009}, so that smaller particles tend to charge negatively and larger ones positively. Whether this differential charging tendency operates one way or the other depends on the microphysics, which differs for different materials, so this polarity differs. For our present purposes, however, what is relevant is that there should exist such a triboelectric charging tendency in one sense or the other. Here we build a minimal model (Fig.~\ref{model}) incorporating solely the collisional dynamics of charge transfer plus nucleation, growth, and fission processes, which we aim to have general relevance to self-replicating granular media.

\begin{figure}[t]
\begin{center}
\includegraphics*[width=0.9\columnwidth]{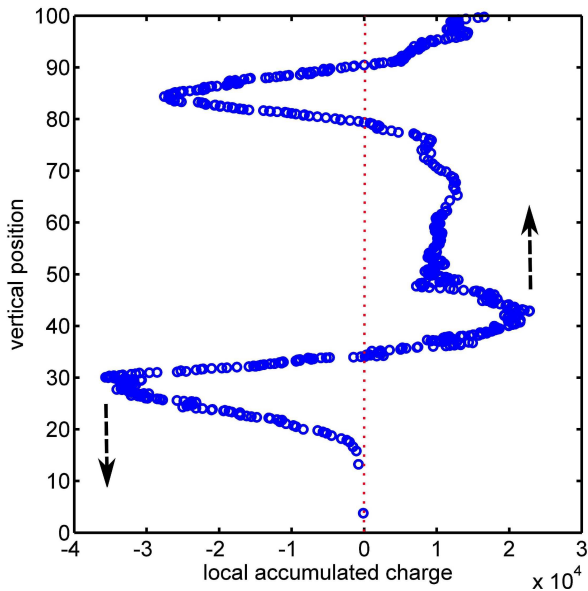}
\end{center}
\caption{\label{chargeseparation} 
Typical transient charge distribution in the model at $t=30$. Charge separation events at two positions are clearly visible. As time passes, light positively charged particles move towards the upper boundary while heavy negatively charged ones move in the opposite direction, as indicated by arrows, contributing to a non-negligible large-scale dipolar charge distribution $\Delta Q$. The neutral charge line (dotted) is marked for reference.}
\end{figure}

We consider (Fig.~\ref{model}) a one-dimensional system of length $L$, within which we place randomly $n$ neutrally charged particles, $q_i=0$ for $i=1,\ldots,n$, of size $r_i$ extracted from a Gaussian distribution of sizes with mean $\bar{r}=1$ and standard deviation $\sigma$. These particles grow at a constant rate, independent of particle size, that sets the timescale of the problem. At the same time, they sediment in an upwards flow of constant magnitude at their terminal velocity, $u_i$, determined by the instantaneous balance of fluid drag and gravitational forces. The sedimentation speed is normalized by the updraft speed (i.e., $u_i=1$ for passive tracers) and is approximated by a linear function of the particle size $r_i$, $u_i=1-(r_i/R_c)$. This function is positive (negative) for $r_i$ below (above) a critical value $R_c$. The sedimentation speed of all particles in the simulation is updated every integration time step; in this way, small particles that are initially advected upward by the updraft slow their upward motion as they grow and begin moving downward after reaching the threshold $R_c$.

When the trajectories of two particles $i$ and $j$ cross, they collide. Both mass and charge are conserved during a collision but, while charge is transferred in every collision (the smaller particle leaving the collision positively charged: $q_i \rightarrow q_i + 1$ and $q_j \rightarrow q_j - 1$ with $r_i < r_j$), mass is only transferred if fission occurs through secondary nucleation. We assign a certain probability $s$ for fission to occur for each of the two particles involved in a collision. If fission does occur for particle $i$, a new, neutrally-charged particle with radius $1$ splits from it, reducing its size to $r_i = r_i-1$. Boundary conditions are absorbing: particles leaving the system through the upper or lower boundary are absorbed there, and do not participate further in the dynamics, but their charge accumulates to the total charge at the boundaries: $Q_{u (l)}=\sum q_i$ for $i$ leaving the system through the upper (lower) boundary, respectively. The total large-scale charge separation is calculated as $\Delta Q = Q_{u} - Q_{l}$.

For given initial conditions (set completely by the concentration of particles, $\rho =n/L$, and the initial spread of the size distribution, $\sigma$), the behavior of our minimal model then depends on just two parameters, the critical radius $R_c$ and the secondary nucleation rate $s$. In what follows we set $Rc=8$, $\rho = 2$ and $\sigma = 0.1$ and explore the behavior of the model with respect to the secondary nucleation rate $s$.

\begin{figure}[t]
\begin{center}
\includegraphics*[width=\columnwidth]{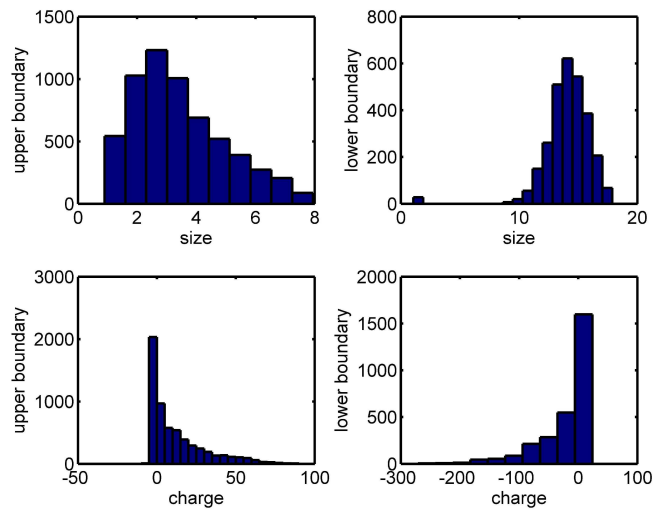}
\end{center}
\caption{\label{hist}
Particle size ($r_j$) and charge ($q_j$) distributions for the system upper boundary (left panels) and lower boundary (right panels). Charge separation, correlated with particle size separation, is observed.}
\end{figure}

Figure~\ref{chargeseparation} displays the typical transient dynamics of the charge distribution in the model. Charge separation events at two vertical positions are clearly visible. The figure shows an initial stage in the transient dynamics chosen to exemplify how the charge separation process operates in the model. As time passes, as indicated by black arrows, light positively charged particles move towards the upper boundary while heavy negatively charged ones move in the opposite direction, contributing to a non-negligible dipolar large scale charge separation $\Delta Q$. Regions where localized collisions are produced generate large charges that are then separated by differential advection. Even at those early stages in the process, the global dipolar structure of the system, quantified by the large-scale charge difference between the boundaries $\Delta Q$, is the dominant field. Figure~\ref{hist} shows particle size and charge distributions for the upper and lower boundaries. Charge separation correlated with particle size separation is observed. 

\begin{figure}[t]
\begin{center}
\includegraphics*[width=0.9\columnwidth]{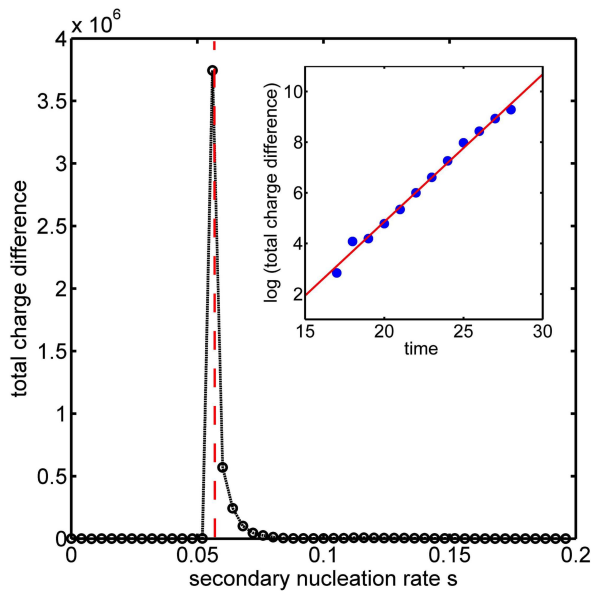}
\end{center}
\caption{\label{secondary} The charge separation shown at a finite time step in the simulation. The total charge separation $\Delta Q$ displays non-monotonic behavior as a function of the secondary nucleation rate $s$. Inset: time evolution of $\Delta Q$ for $s=0.1$ shows the exponential growth characteristic of all values of $s>s_c$.}
\end{figure}

We display the dynamics of the total charge separation $\Delta Q$ in Fig.~\ref{secondary}. The total charge separation $\Delta Q$ at a finite time step (which is a measure of the speed of the charge-separation process) displays non-monotonic discontinuous behavior as a function of the secondary nucleation rate $s$. Below a critical value $s_c$, the long-term dynamics reaches an emptying state with no particles left in the domain and a finite (and small) final charge separation; as there are few secondary nuclei, the initial particles grow and collide very little before leaving the system. Occasionally some charge is produced, but not much. Above $s_c$ the number of collisions grows, and with it the charge transfer. As the charge acquired by the particles is correlated with their size, charge separation occurs too. The system approaches a steady state with an average non-zero number of particles in the domain and an exponentially divergent charge separation, as shown in the inset of Fig.~\ref{secondary}. This exponential charge growth is limited in nature by electrical discharges; `lightning'. 

\begin{figure*}[t]
\begin{center}
\includegraphics*[width=0.9\textwidth]{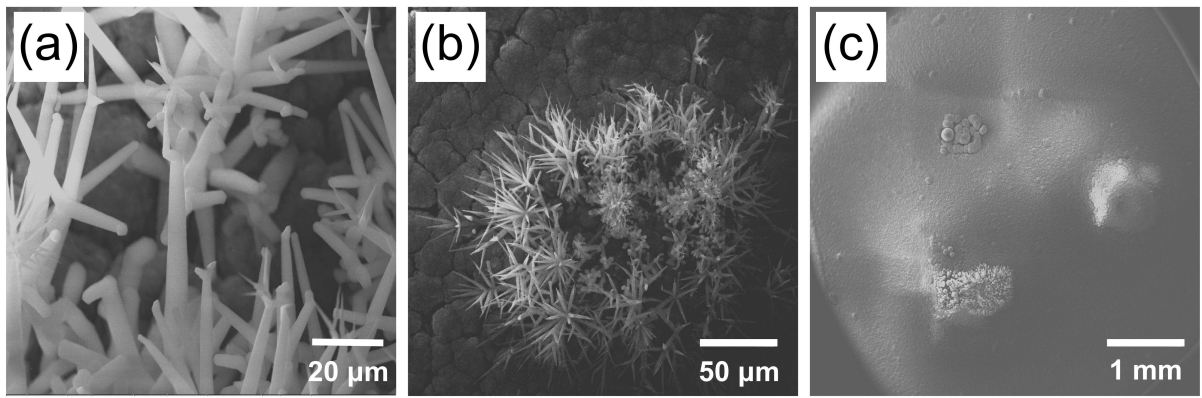}
\end{center}
\caption{\label{charging}
An electric field promotes ice `palm' growth:
(a, b) At $T = 170$~K, $P = 40$~Pa,  $V=30$~kV, an ice `forest' is rapidly formed. The forest displays the morphology intermediate between whiskers and dendrites typically formed with charged ice.
(c) In an overview the three zones where we imaged and charged with electrons stand out, showing the increased growth on the background ice film.
}
\end{figure*}

However, for secondary-nucleation rates much greater than $s_c$, there are more and more collisions, the particles undergo more fission into secondary nuclei, and, although much charge is produced, it does not separate as well; there is thus an intermediate optimal value of the secondary nucleation rate to produce the greatest charge separation. This critical value $s_c \approx  1/(\rho Rc)$ is $0.0625$ in the simulations (dashed line in Fig.~\ref{secondary}). A smaller value of $R_c$ (which means a weaker updraft) requires secondary nucleation to occur more often (larger $s_c$) in order to ensure some particles are advected upwards (leading to charge separation), while the initial density of particles, $\rho$, controls the collision probability. It is worth mentioning that even in the limit of very high particle density $\rho$, in which many collisions occur at early stages, charge separation is minimal in the absence of secondary nucleation. It is the latter that is responsible for the observed discontinuous critical behavior.

Earlier work of ours made us suspect that secondary nucleation ought to be important in runaway collisional charging. Previously we have shown that the nonlinear feedback effects of secondary nucleation are responsible for chiral symmetry breaking in experiments involving crystallizing a chiral chemical compound from solution \cite{chiral1,chiral2}. We showed that secondary nuclei in such stirred crystallization experiments are often easily-detached whisker or needle crystallites growing from a mother crystal, and a runaway process involving the formation of secondary nuclei leads to complete chiral symmetry breaking.

Water molecules possess a high electric polarizability; they are electrical dipoles and can be highly affected by the presence of an external field. During dendrite growth, an electric field can produce an ordering of the molecular dipoles and increase the molecular flow towards the dendrite tip. This increases the growth velocity, decreases the tip radius and disables the generation of side-branches, producing long whiskers. These effects have long been noted \cite{mason} and were studied quantitatively by Libbrecht and Tanusheva \cite{libbrecht}, who measured the tip velocity and found that high voltages could multiply the growth rate more than tenfold. 

We hypothesized that this dendrite growth mechanism should be involved in promoting secondary nucleation. Thus we undertook laboratory experiments to see whether similarly easily-detached forms as in solution crystallization experiments are produced in ice under the influence of an electric field.  We employed a FEI Quanta 200 environmental scanning electron microscope (ESEM) equipped with a liquid nitrogen cold stage to grow ice {\em in situ} at low pressures and at temperatures of 90--200~K. The microscope was set up so that the cold finger, together with a thermostat, was directly beneath the substrate (a silicon wafer attached with silver glue). We began by evacuating the chamber in the high-vacuum mode of the microscope (6 x $10^{-4}$ Pa) and lowering the substrate to the working temperature. We first scanned the uncovered sample substrate, on which we grew an ice film by switching to low-vacuum mode and opening the water input microvalve at a pressure of 40~Pa for some seconds.  We found this was the highest pressure at which we could obtain clear images. We closed the microvalve at or before the point when the substrate temperature increases and cannot be maintained at the working temperature, following which we switched back to high-vacuum mode and observed the ice growth \emph{in situ}.

In the electron-microscope chamber a high-voltage electron beam is used for imaging and, as we display in Fig.~\ref{charging}, we find that charging effects produce rapid ice growth wherever we charge with electrons by imaging. A typical ice morphology seen under these circumstances is of a form intermediate between whiskers and dendrites, which often takes on the aspect of a palm tree; Fig.~\ref{charging}a. As we zoom out,  (Fig.~\ref{charging}b, c), we note that the palm forest is found only where we had been imaging; outside the area of the electron beam, we find a relatively flat film of ice; while within the imaged zones --- three in Fig.~\ref{charging}c --- we find faster ice growth and the ice forest. 

We had noted such palm-like forms in previous experiments involving growing ice inside an electron microscope \cite{astrophys_j}, but had not then realized that the electric field was involved in their production. These experiments are necessarily qualititative, being performed within the chamber of an unmodified ESEM, but we find the results suggestive: Owing to their geometry, the breakage of these structures on collision is likely and will lead to the formation of new nucleation centres. Such friable morphologies do not only form in ice under electric fields; snowflakes too have such delicate structures, but an electric field promotes this form of growth \cite{libbrecht}.

Our minimal physical model of a self-replicating granular material shows how secondary nuclei from
such growth can lead to runaway charging. These effects may be present in ice on Earth, in terrestrial thunderstorms \cite{vonnegut}, and in astrophysical ices, in the solar nebula \cite{gibbard,pilipp,muranushi} and in thunderstorms on other planets  \cite{aplin,yair}, some of which, e.g., on Venus, may involve self-replicating granular materials other than water ice. It is conceivable that this dynamics is involved in the formation of the Martian geological structures called razorbacks \cite{Shinbrot:2006}.

We thank Werner Kuhs and Emmanuel Villermaux for useful discussions and Karin Whitmore for her assistance in the electron microscopy sessions. We acknowledge the financial support of the Spanish MICINN grant FIS2010-22322-C01 (OP and IT), ``subprograma Ram\'on y Cajal'' (IT), and grant FIS2010-22322-C02 (JHEC and CISD); and the Austrian Science Fund (FWF), grant P23027 (HG).


\end{document}